\documentclass[cameraready]{Interspeech}

\title{Beyond Semantic Dominance: Cognitive Affective Reasoning and Empathetic Response Alignment in Audio Language Models}

\author[affiliation={1}, equalcontribution]{Zhixian}{Zhao}
\author[affiliation={1}, equalcontribution]{Shuiyuan}{Wang}
\author[affiliation={1}]{Wenjie}{Tian}
\author[affiliation={1}]{Jingbin}{Hu}
\author[affiliation={1}]{Ziyu}{Zhang}
\author[affiliation={1},correspondingauthor]{Lei}{Xie}
\address{
    $^1$ Audio, Speech and Language Processing Group (ASLP@NPU),
    Northwestern Polytechnical University, Xi'an, China
}

\email{zxzhao@mail.nwpu.edu.cn, lxie@nwpu.edu.cn}


\usepackage{comment}
\usepackage{booktabs}
\usepackage{multirow}
\usepackage{graphicx}

\begin{document}

\maketitle

\begin{abstract}
While Audio Language Models (ALMs) demonstrate strong semantic understanding, they struggle with complex affective interactions. Specifically, textual semantic dominance often overshadows acoustic nuances, and a lack of cognitive depth leads to generic, emotion-agnostic responses. We propose CogAudio-LLM\footnote{ \urlstyle{same} \url{https://github.com/zxzhao0/CogAudio-LLM}}, a novel cognitive affective reasoning framework. To mitigate semantic dominance, we build LIME-440K, a ``lexically-identical, multi-emotion'' dataset designed to facilitate acoustic-semantic decoupling. We introduce EIPS, a 4-step Chain-of-Thought (CoT) mechanism incorporating psychological reasoning. For inference efficiency, multi-stage training explicitly establishes EIPS via supervised fine-tuning, then distills this logic into an implicit generation process. Finally, we design DR-SAPO (Dual-Route Soft Adaptive Policy Optimization) to dynamically balance the logical rigor of the CoT with the empathetic quality of the direct response.

\end{abstract}

\vspace{-0.2cm}
\section{Introduction}

By extending Large Language Models (LLMs) into the audio modality, Audio Language Models (ALMs) have significantly advanced natural spoken dialogue interaction~\cite{osum,e-chat,emova}. Specifically, to meet the strong demands for emotional companionship in real-world scenarios, recent research primarily follows two trajectories: general-purpose speech foundation models that treat emotion understanding as a downstream evaluation task~\cite{qwen-audio,qwen2audio,salmonn,kimi}, and empathetic dialogue systems that leverage audio perception to generate contextually emotionally-appropriate responses~\cite{osum,emoomni}. However, achieving true empathy requires models to not only accurately perceive fine-grained paralinguistic cues but also respond with appropriate empathetic alignment.

\begin{figure}[t]
    \centering
    \includegraphics[width=\linewidth]{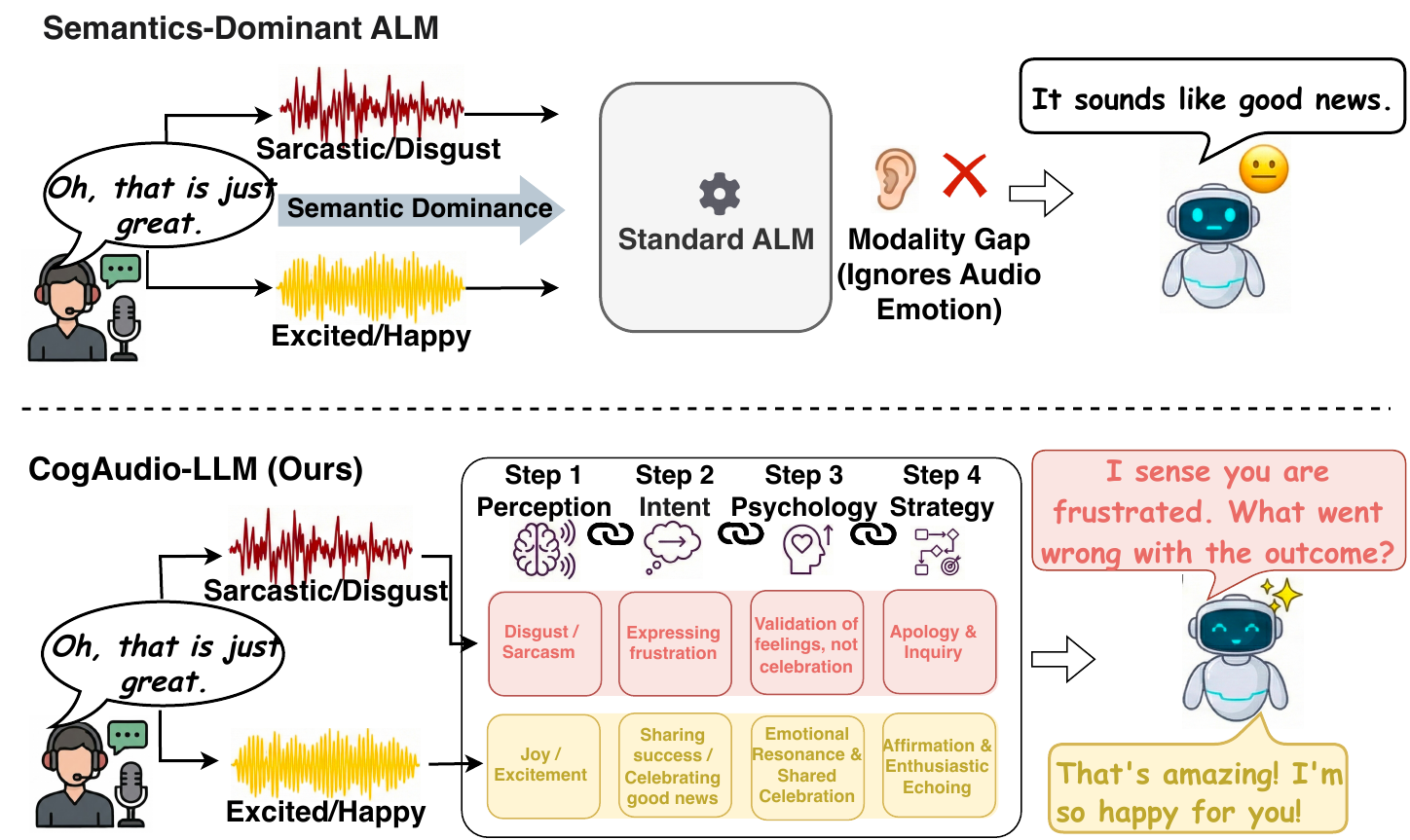}
    \caption{Standard ALMs misjudge emotions due to semantic dominance and the modality gap. In contrast, our CogAudio-LLM uses a 4-step EIPS cognitive framework to ensure precise empathetic alignment.}
    \vspace{-0.3cm} 
    \label{fig:concept}
    \vspace{-0.3cm}
\end{figure}

Current ALMs still face two critical bottlenecks in deep affective interaction. The first is the phenomenon of \textit{Semantic Dominance}. Fundamentally, most ALMs are built upon pre-trained text-native LLMs. Due to the massive scale of text pre-training, their latent representation spaces are inherently biased towards discrete textual semantics~\cite{text1,text2}. Consequently, in multimodal processing, semantic representations often overshadow acoustic details. As a result, when acoustic cues contradict literal textual semantics (e.g., in scenarios of sarcasm or forced smiles), models exhibit a modality gap~\cite{conflict1,conflict2,SABER}. 
Relying excessively on text, they misjudge the user's true emotional state, leading to inappropriate responses, as shown in the upper part of Fig.~\ref{fig:concept}. Second, they lack \textit{Cognitive Depth}. Even when emotions are correctly identified, standard ALMs tend to generate generic, context-agnostic responses (e.g., ``safe'' empathy templates). Although recent studies in the speech domain~\cite{c2ser,emotionthinker} have successfully introduced Chain-of-Thought (CoT) into ALMs to enhance emotion understanding by explicitly prompting models to describe acoustic features (e.g., pitch and speaking rate) prior to emotion prediction, these approaches still rely on acoustic-descriptive reasoning. Consequently, they struggle to reliably infer user's latent intents and true psychological states in complex interactions.

To bridge these gaps, we propose \textit{CogAudio-LLM}, a cognitive emotional reasoning framework for spoken interaction. To overcome semantic bias, we construct LIME-440K (``Lexically-Identical, Multi-Emotion''), a large-scale bilingual dataset containing approximately 440,000 utterances. This design encourages the model to decouple acoustics from semantics, strictly relying on paralinguistic cues for fine-grained emotion perception. Addressing the lack of cognitive depth, we introduce EIPS (Emotion Perception, Intent Extraction, Psychological Modeling, and Strategy Formulation), a 4-step CoT reasoning mechanism. This explicit structure embeds rigorous psychological logic into the response generation process. 
Furthermore, to enable the model to generate highly empathetic responses as intuitively as humans, we adopt a multi-stage training paradigm. We first establish explicit EIPS reasoning capabilities via Supervised Fine-Tuning (SFT) and then implicitly internalize these capabilities using mixed response-only data. Finally, we introduce DR-SAPO (Dual-Route Soft Adaptive Policy Optimization) to dynamically reward either the logical rigor of the CoT or the empathy depth of the direct response. To facilitate future research, we release the LIME-440K dataset.

The main contributions of this paper as follows:
\begin{itemize}
    \item We propose CogAudio-LLM, a novel cognitive emotional reasoning framework that significantly enhances the emotional insight and empathetic response capabilities of ALMs in complex interactions by integrating an explicit EIPS psychological Chain-of-Thought.
    \item We construct and release LIME-440K, a large-scale bilingual dataset explicitly designed with a semantic-acoustic decoupling strategy. This resource mitigates the `semantic dominance' bottleneck and improves fine-grained acoustic emotion perception.
    \item We design a multi-stage internalization training paradigm combined with the DR-SAPO dual-route reinforcement learning algorithm, achieving the implicit internalization of reasoning capabilities while dynamically aligning logical rigor and empathetic depth.
    \item Experimental results demonstrate that CogAudio-LLM significantly outperforms existing state-of-the-art baselines in both fine-grained emotion recognition accuracy and the quality of empathetic alignment.
\end{itemize}

\begin{figure*}[t]
    \centering
    \includegraphics[width=\textwidth]{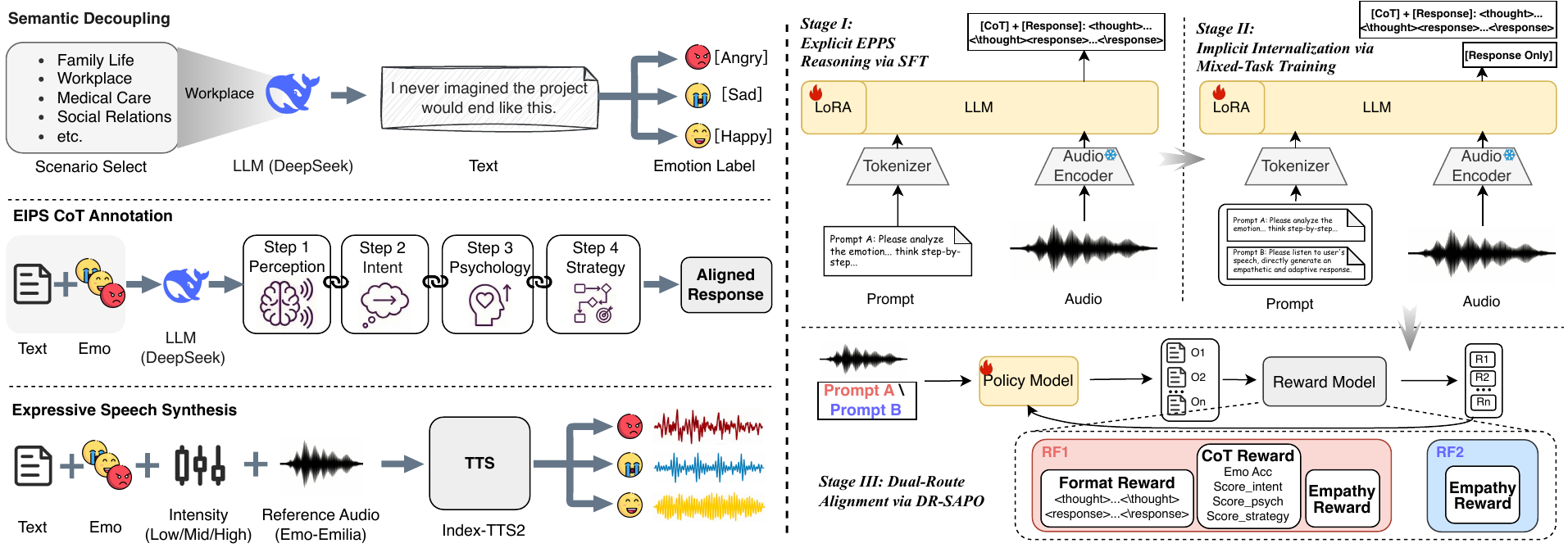}
    \vspace{-0.2cm} 
    \caption{The pipeline of data generation and model training. \textbf{Left:} Construction of the LIME-440K dataset via semantic decoupling, EIPS CoT annotation, and expressive synthesis. \textbf{Right:} The three-stage training architecture of CogAudio-LLM, encompassing explicit reasoning SFT, implicit mixed-task internalization, and DR-SAPO dual-route alignment.}
    \label{fig-pipeline}
    \vspace{-0.6cm} 
\end{figure*}

\vspace{-0.3cm}
\section{LIME-440K Dataset}

To address the fundamental limitations of existing speech datasets, where textual semantics and acoustic emotions are highly coupled and explicit reasoning paths are lacking, we construct LIME-440K (\textbf{L}exically-\textbf{I}dentical, \textbf{M}ulti-\textbf{E}motion), a large-scale bilingual dataset. As shown in Table~\ref{tab:lime_stats}, this dataset comprises approximately 440,000 speech utterances totaling roughly 497 hours. To balance emotion specificity and data distribution diversity, the dataset consists of two subsets:
\begin{itemize}
    \item \textbf{LIME-Core (Part A and B)}: A core subset covering 7 fine-grained emotions, constructed specifically using our semantic-acoustic decoupling strategy.
    \item \textbf{LIME-Aug (Part C and D)}: An augmented subset that introduces and re-annotates open-source data (specifically, ECD-TSE~\cite{ecd-tse} and ESD~\cite{ESD}) to expand speaker voices and mapping patterns, thereby enhancing model generalization.
\end{itemize}
As illustrated in the left part of Fig.~\ref{fig-pipeline}, the construction of LIME-Core primarily involves the following three stages:

\vspace{-0.2cm}
\subsection{Semantic-Acoustic Decoupled Generation}
To break the model's over-reliance on text (i.e., ``Semantic Dominance''), we adopt a ``one-text, multi-emotion'' data generation paradigm. Following~\cite{scene}, we pre-define 20 core interaction scenarios covering family, work, and other domains. In each scenario, we utilize DeepSeek-V3~\cite{deepseek-v3} model to generate text with high semantic ambiguity, ensuring that the same text must adapt to at least three distinctly different emotion labels. For example, the text ``I never imagined the project would end like this'' must be paired with contextual settings for [Happy], [Sad], and [Angry]. This design effectively mitigates the direct text-to-emotion mapping shortcut, compelling the model to disambiguate strictly via paralinguistic cues.

\vspace{-0.2cm}
\subsection{EIPS Chain-of-Thought Annotation}
To incorporate structured psychological reasoning paths into the dataset, we leverage the DeepSeek-R1~\cite{deepseekr1} model for knowledge distillation. We design structured prompts guiding the model to take a given ``text + emotion label'' as input and automatically generate a Chain-of-Thought (CoT) that strictly follows the four steps of EIPS, along with the final aligned response. These four steps include:
\begin{itemize}
    \item \textbf{Emotion Perception}: Parse explicit/implicit emotional elements, evaluate emotion intensity, and locate trigger points.
    \item \textbf{Intent Extraction}: Strip away superficial utterances to unearth the user's unmet deep psychological needs.
    \item \textbf{Psychological Modeling}: Anticipate the interlocutor's cognitive biases and defense mechanisms, and plan the emotional landing point.
    \item \textbf{Strategy Formulation}: Design a dialogue path conforming to emotional development patterns and a response scheme adapted to the cultural context.
\end{itemize}

To ensure annotation quality, 400 random samples were evaluated by annotators with linguistics backgrounds, achieving a 93\% acceptance rate based on logical coherence and emotional accuracy.

\vspace{-0.2cm}
\subsection{Expressive Speech Synthesis}
\vspace{-0.2cm}
To ensure the generated speech highly aligns with the fine-grained emotion labels, we employ the Index-TTS2~\cite{index-tts2} model for high-fidelity speech synthesis. During this process, alongside text and emotion categories, we introduce ``emotion intensity'' (low/mid/high) as an independent control variable. This design aims to capture intra-class acoustic variance within the same emotion category, preventing the model from forming singular acoustic mappings (e.g., restricting ``Angry'' solely to high-volume outbursts). Such fine-grained control parameters help improve the model's perception of complex emotional expressions. Furthermore, to bridge the acoustic gap between synthesized speech and real-world scenarios, we introduce utterances containing real environmental noise from the Emo-Emilia dataset as reference audio, further enhancing the model's robustness in open acoustic environments.

\begin{table}[t]
    \centering
    \caption{Statistics of the LIME-440K Dataset.}
    \vspace{-0.3cm} 
    \setlength{\tabcolsep}{4pt} 
    \resizebox{\columnwidth}{!}{%
        \begin{tabular}{l c c c c r}
            \toprule
            \textbf{Subset} & \textbf{Lang.} & \textbf{Emo.$\times$Int.} & \textbf{Spk.} & \textbf{Hrs} & \textbf{Utt.} \\
            \midrule
            \multicolumn{6}{l}{\textit{\textbf{LIME-Core}}} \\
            \hspace{0.5em}Part A (CN) & CN & 7$\times$3 & $\sim$200 & 263.9 & 223,884 \\
            \hspace{0.5em}Part B (EN) & EN & 7$\times$3 & $\sim$200 & 113.8 & 96,000 \\
            \midrule
            \multicolumn{6}{l}{\textit{\textbf{LIME-Aug}}} \\
            \hspace{0.5em}Part C (ECD-TSE) & EN & 5$^{\dagger}$$\times$1 & 12 & 90.3 & 84,000 \\
            \hspace{0.5em}Part D (ESD) & Mix & 5$^{\ddagger}$$\times$1 & 20 & 29.1 & 35,000 \\
            \midrule
            \textbf{Total} & \textbf{CN/EN} & \textbf{7} & \textbf{$\sim$230} & \textbf{497.1} & \textbf{438,884} \\
            \bottomrule
            \multicolumn{6}{l}{\scriptsize $^{\dagger}$Excludes Disgusted/Surprised. $^{\ddagger}$Excludes Fear/Disgusted. Int.: 3 levels.}
        \end{tabular}%
    }
    \label{tab:lime_stats}
    \vspace{-0.5cm}
\end{table}

\vspace{-0.2cm}
\section{CogAudio-LLM}
\label{sec:training}
To balance detailed affective reasoning with fluid conversational interaction, we propose a three-stage ``explicit guidance and implicit internalization'' training paradigm, as shown in Fig.~\ref{fig-pipeline}. 
Furthermore, we design a Dual-Route Soft Adaptive Policy Optimization (DR-SAPO) algorithm to align complex emotional cognition with human preferences. 

\vspace{-0.2cm}
\subsection{Stage I: Explicit EIPS Reasoning via SFT}

First, using Qwen2.5-omni~\cite{Qwen2.5-Omni} as our foundational model, we establish structured psychological reasoning capabilities.
In this stage, we utilize the core subset of LIME-440K containing comprehensive EIPS annotations. Using Prompt A (``Please analyze the emotion... think step-by-step''), we instruct the model to sequentially generate the full 4-step EIPS Chain-of-Thought (CoT) prior to outputting the final response for a given audio input $X_a$. 
Given the target sequence $Y = [Y_{CoT}, Y_{res}]$, the optimization objective is the standard auto-regressive negative log-likelihood loss:
\vspace{-0.2cm}
\begin{equation}
\label{eq:stage1}
L_{Stage1} = - \sum_{t=1}^{|Y|} \log P_{\theta}(y_t | X_a, P_A, y_{<t}).
\end{equation}
\vspace{-0.3cm}

Through Stage I, the model establishes an explicit mapping logic from acoustic perception to deep psychological strategic planning.

\subsection{Stage II: Implicit Internalization via Mixed-Task Training}
While explicit CoT provides deep psychological analysis, relying on it for every conversational turn diverges from natural, intuitive human interaction. To achieve the ``implicit internalization'' of reasoning capabilities, Stage II incorporates data containing only direct responses (without CoT), triggered by Prompt B (``Please listen to user's speech, directly generate an empathetic response''). 
We construct a mixed distribution $\mathcal{D}_{mix}$ by combining explicit reasoning data (Prompt A) and implicit response data (Prompt B) at a 1:1 sampling ratio for joint training:
\vspace{-0.2cm}
\begin{equation}
\label{eq:stage2}
L_{Stage2} = - \mathbb{E}_{(X_a, P_i, Y_i) \sim \mathcal{D}_{mix}} \left[ \sum_{t=1}^{|Y_i|} \log P_{\theta}(y_t | X_a, P_i, y_{<t}) \right],
\end{equation}
where $P_i \in \{P_A, P_B\}$. This mixed mechanism with shared underlying parameters allows the model, when executing the pure response task (skipping intermediate outputs), to implicitly activate the EIPS cognitive circuits learned in Stage I, thereby generating responses of equivalent empathetic depth.

\subsection{Stage III: Dual-Route Alignment via DR-SAPO}
To further enhance the logical rigor of emotional reasoning and the empathetic depth of the final response, we introduce Reinforcement Learning (RL) in the third stage. 
To address gradient instability and learning signal loss common in long-sequence CoT generation, we employ the advanced Soft Adaptive Policy Optimization (SAPO) algorithm~\cite{sapo} as our RL foundation. SAPO replaces the hard clipping of traditional PPO~\cite{PPO}/GRPO~\cite{GRPO} with a smooth soft-gating mechanism, significantly improving training stability for long sequences.

Building upon this foundation, we propose the core DR-SAPO mechanism. For an input query $q$ and generated output $o_k$, DR-SAPO dynamically allocates a differentiated dual-route reward $R(o_k)$ based on the triggered prompt:
\vspace{-0.2cm}
\begin{equation}
\label{eq:drsapo}
R(o_k) = \mathbb{I}_{\{P_A \in q\}} \cdot RF_1(o_k) + \mathbb{I}_{\{P_B \in q\}} \cdot RF_2(o_k),
\end{equation}
where $\mathbb{I}$ represents the indicator function. Crucially, both routes share a Response Empathy Reward ($R_{res}$), evaluated by Gemini2.5-pro~\cite{gemini2.5} acting as an LLM-as-a-Judge based on acoustic emotion specificity and psychological insight. This metric strictly penalizes generic templates detached from acoustic features and heavily rewards responses that precisely address hidden psychological needs. Consequently, the distinct dual-route allocation is defined as follows:

\begin{itemize}
    \item \textbf{Route 1 (Explicit Reasoning $RF_1$):} When CoT-inclusive reasoning ($P_A$) is triggered, the system evaluates the entire reasoning-to-response pipeline. To explicitly balance the multi-dimensional evaluation, $RF_1$ is formulated as a weighted linear combination:
    \vspace{-0.15cm}
    \begin{equation}
    \label{eq:rf1}
    RF_1(o_k) = \lambda_{fmt} R_{fmt} + \sum_{c \in \mathcal{C}} \lambda_c R_{c} + \lambda_{res} R_{res},
    \end{equation}
    where $R_{fmt}$ is the format reward (ensuring correct closure of \texttt{<thought>} and \texttt{<response>} tags), and $\mathcal{C} = \{emo, intent, psych, strategy\}$ represents the set of fine-grained CoT logic rewards evaluated by the LLM-as-a-Judge. The hyper-parameters $\lambda$ dictate the relative importance of each cognitive dimension during RL optimization.
    \item \textbf{Route 2 (Implicit Response $RF_2$):} When a direct response ($P_B$) is triggered, the model omits the CoT. Here, $RF_2$ consists solely of the empathy reward $R_{res}$.
\end{itemize}

Through this differentiated reward design, Route 1 encourages logical rigor in psychological deduction, while Route 2 optimizes the model to maintain empathetic alignment when explicit reasoning is bypassed. This facilitates the implicit internalization of affective cognition.

\vspace{-0.2cm}
\section{Experiments and Results}
\label{sec:experiments}

\subsection{Experimental Setup}
\vspace{-0.2cm}
\textbf{Datasets and Evaluation:} We train on LIME-440K and evaluate on two distinct benchmarks: (1) ESD-Test contains 1000 real human utterances from 2 strictly held-out speakers (5 emotions) to evaluate zero-shot speaker generalization; and (2) Humdial-EIBench\footnote{ \urlstyle{same}  \url{https://huggingface.co/datasets/ASLP-lab/HumDial-EIBench}}~\cite{HumDial-EIBench,humdial} Task4 comprises 200 bilingual utterances of real-world, spontaneous human speech (7 emotions). To rigorously test semantic decoupling, this dataset is evenly split into semantic-acoustic consistent and conflicting (e.g., sarcasm) subsets. Our evaluation encompasses two dimensions: (1) \textbf{Emo-Acc}, an objective metric measuring fine-grained emotion classification accuracy; and (2) \textbf{Empathy Quality}, a subjective 1-4 point scale (1: generic templates, 4: specific deep psychological insights. Detailed 1-4 scoring rubrics are provided on our repository). To ensure fairness, this metric is cross-validated by Gemini2.5-pro~\cite{gemini2.5} acting as an LLM-as-a-Judge and five independent human experts (ICC = 0.78, indicating good inter-rater reliability).

\textbf{Baselines:} We compare CogAudio-LLM against cutting-edge open-source models: Freeze-Omni~\cite{freeze-omni}, GLM-4-Voice~\cite{glm4}, Kimi-Audio~\cite{kimi}, Step-Audio-2-mini~\cite{step-audio2}, Qwen2.5-Omni-7B~\cite{Qwen2.5-Omni}, Qwen3-Omni-30B~\cite{qwen3-omni}, and the commercial API GPT-4o-audio~\cite{gpt4o}.

\textbf{Implementation Details:} 
CogAudio-LLM is initialized with Qwen2.5-omni-7B~\cite{Qwen2.5-Omni} and trained using 8 NVIDIA A100 GPUs. Both SFT stages (Stage I and II) utilize LoRA (rank $r=8$, $\alpha=32$) with a learning rate of 1e-5, a batch size of 512, and are trained for 3 epochs. The DR-SAPO stage updates for 1500 steps with a reduced learning rate of 1e-6 and a batch size of 64. Based on empirical tuning, the Route 1 reward hyper-parameters defined in Eq.~\ref{eq:rf1} are configured as follows: $\lambda_{fmt} = 0.1$, $\lambda_{res} = 0.3$, and for the CoT components, $\lambda_{emo} = 0.3$, $\lambda_{intent} = 0.1$, $\lambda_{psych} = 0.1$, and $\lambda_{strategy} = 0.1$, where $\lambda_{emo}$ is prioritized to ensure accurate emotional anchoring for downstream reasoning.

\begin{table}[t]
    \centering
    \caption{Comparison of Empathy Quality evaluated by LLM and Human (scored on a 1-4 scale). CogAudio-LLM utilizes implicit responses.}
    \vspace{-0.3cm} 
    \label{tab:main_results}
    \setlength{\tabcolsep}{3.5pt} 
    \resizebox{\columnwidth}{!}{%
        \begin{tabular}{l c c c c c}
            \toprule
            \multirow{2}{*}{\textbf{Model}} & \multirow{2}{*}{\textbf{LLM (ESD)}} & \multicolumn{2}{c}{\textbf{LLM (HumDial)}} & \multicolumn{2}{c}{\textbf{Human (HumDial)}} \\
            \cmidrule(lr){3-4} \cmidrule(lr){5-6}
            & & \textbf{Conf.} & \textbf{Non-conf.} & \textbf{Conf.} & \textbf{Non-conf.} \\
            \midrule
            Freeze-Omni          & 1.34 & 1.34 & 2.12 & 1.56 & 1.90 \\
            GLM-4-Voice          & 1.42 & 1.79 & 2.09 & 1.67 & 2.21 \\
            Kimi-Audio           & 1.54 & 1.53 & 2.16 & 1.90 & 2.29 \\
            Step-Audio-2-mini    & 1.22 & 1.58 & 1.95 & 1.89 & 2.11 \\
            Qwen2.5-Omni-7B      & 1.64 & 1.75 & 2.40 & 2.14 & 2.51 \\
            Qwen3-Omni-30B       & 1.52 & 1.86 & 2.38 & 1.78 & 2.01 \\
            GPT-4o-Audio         & 1.59 & 1.82 & 2.58 & 1.68 & 2.05 \\
            \midrule
            \textbf{CogAudio-LLM} & \textbf{2.90} & \textbf{2.91} & \textbf{3.24} & \textbf{3.16} & \textbf{3.17} \\
            \bottomrule
        \end{tabular}%
    }
    \vspace{-0.2cm}
\end{table}

\begin{table}[t]
\centering
\caption{Ablation study on Emotion Perception Accuracy (\%). `-' indicates that the metric is not applicable for the specific training stage.}
\vspace{-0.3cm} 
\label{tab:ablation_emo}
\resizebox{\linewidth}{!}{
\begin{tabular}{l ccc}
\toprule
\multirow{2}{*}{\textbf{Model / Training Stage}} & \multicolumn{3}{c}{\textbf{Emotion Acc. (\%)}} \\
\cmidrule(lr){2-4}
 & \textbf{ESD} & \textbf{Conflict} & \textbf{Non-conf} \\
\midrule
\textbf{Qwen2.5-omni} (Base) & 26.5 & 24.0 & 68.0 \\
\textbf{A. Base (Direct SFT)} & - & - & - \\
\textbf{B. Explicit Only SFT} & 47.5 & 42.0 & \textbf{73.0} \\
\textbf{C. Ours w/o RL} & 47.0 & 44.0 & \textbf{73.0} \\
\textbf{D. Ours (Full) w/ RL} & \textbf{49.5} & \textbf{46.0} & 71.0 \\
\bottomrule
\end{tabular}
}
\vspace{-0.3cm}
\end{table}

\begin{table}[t]
\centering
\caption{Ablation study on Empathy Quality (1-4 scale) across different test sets. Impl. and Expl. denote the Implicit response and Explicit CoT reasoning routes. `-' indicates not applicable.}
\vspace{-0.3cm} 
\label{tab:ablation_empathy}
\setlength{\tabcolsep}{3pt} 
\resizebox{\linewidth}{!}{
\begin{tabular}{l ccc ccc}
\toprule
\multirow{2}{*}{\textbf{Model / Training Stage}} & \multicolumn{3}{c}{\textbf{Implicit Response}} & \multicolumn{3}{c}{\textbf{Explicit CoT}} \\
\cmidrule(lr){2-4} \cmidrule(lr){5-7}
 & \textbf{ESD} & \textbf{Conflict} & \textbf{Non-conf} & \textbf{ESD} & \textbf{Conflict} & \textbf{Non-conf} \\
\midrule
\textbf{Qwen2.5-omni} (Base)  & 1.64 & 1.75 & 2.40 & 1.54 & 1.64 & 2.64 \\
\textbf{A. Base (Direct SFT)} & 2.10 & 2.62 & 3.11 & - & - & - \\
\textbf{B. Explicit Only SFT} & - & - & - & 2.39 & 2.35 & 3.16 \\
\textbf{C. Ours w/o RL}       & 2.26 & 2.61 & 3.09 & 2.43 & 2.71 & 3.24 \\
\textbf{D. Ours (Full) w/ RL} & \textbf{2.90} & \textbf{2.91} & \textbf{3.24} & \textbf{2.92} & \textbf{2.89} & \textbf{3.39} \\
\bottomrule
\end{tabular}
}
\vspace{-0.5cm}
\end{table}


\vspace{-0.2cm}
\subsection{Main Results}

As shown in Table~\ref{tab:main_results}, we evaluate the subjective empathy quality of all models in the implicit (direct response) mode. Experimental results demonstrate that CogAudio-LLM significantly outperforms all state-of-the-art open-source baselines and the closed-source GPT-4o-Audio across all test sets, which is consistently verified by both LLM and human expert evaluations. 

Most notably, in the semantic-acoustic ``Conflict'' subset, where the literal text contradicts the actual acoustic emotion (e.g., sarcasm or forced smiles), existing advanced models exhibit a severe modality gap, with their empathy scores mostly falling below 2.0 across both evaluation metrics. This indicates that they suffer from the ``Semantic Dominance'' bias, misclassifying the user's true emotion and generating contradictory responses. In contrast, CogAudio-LLM achieves scores of 2.91 (LLM) and 3.16 (Human) on the conflict set, outperforming the baselines by a notable margin. This demonstrates that our framework effectively prioritizes paralinguistic cues over textual semantic interference, enabling the generation of contextually appropriate empathetic responses during conflicting interactions.

\vspace{-0.2cm}
\subsection{Ablation Study}

We conduct a comprehensive ablation study to investigate the contribution of each training stage.

\noindent\textbf{Improvements in Emotion Perception (Table~\ref{tab:ablation_emo}):} The foundational Qwen2.5-omni achieves only 24.0\% accuracy on the ``Conflict'' set, demonstrating a heavy reliance on semantic bias. After explicit fine-tuning on our decoupled LIME-440K (Models B-D), this accuracy nearly doubles (up to 46.0\%). This confirms that our paradigm effectively severs the text-to-emotion shortcut, promoting robust acoustic emotion perception.

\noindent\textbf{Empathy Internalization (Table~\ref{tab:ablation_empathy}):} Direct fine-tuning (Model A) lacks cognitive depth, yielding templated responses. Introducing the EIPS CoT (Model C) structurally enhances empathy. Crucially, our mixed-task training enables Model C's implicit response (2.61) to rival its explicit reasoning (2.71), marking successful ``implicit internalization.'' 
Finally, the DR-SAPO dual-route RL (Model D, i.e., our complete CogAudio-LLM) further aligns the model's implicit reasoning, increasing the empathy score in conflict scenarios to 2.91 and effectively bridging accurate perception with empathetic response generation.

\vspace{-0.2cm}
\section{Conclusion}

We propose CogAudio-LLM, a framework designed to mitigate the ``semantic dominance'' bias in ALMs and enhance cognitive depth during affective interactions. By leveraging the large-scale LIME-440K dataset, embedding the EIPS Chain-of-Thought, and applying the DR-SAPO algorithm, our framework effectively incorporates and streamlines empathetic reasoning. Although CogAudio-LLM demonstrates robust zero-shot generalization to real human speech (shown by ESD and HumDial), a subtle gap remains between TTS training data and spontaneous micro-prosody. Future work will explore mixing in-the-wild datasets during SFT to close this gap. Extensive experiments demonstrate that CogAudio-LLM achieves state-of-the-art performance in fine-grained emotion recognition and empathetic response generation, enabling more context-aware and empathetically aligned spoken dialogue systems.

\section{Generative AI Use Disclosure}
Generative AI models, including DeepSeek-V3/R1, Index-TTS2, and Gemini 2.5 Pro, were used for data generation, EIPS CoT annotation, and response evaluation. The authors are fully responsible and accountable for the final content of this paper. All authors agree with the submission of this paper.

\bibliographystyle{IEEEtran}
\bibliography{mybib}

@inproceedings{ecd-tse,
  author       = {Rui Liu and
                  Pu Gao and
                  Jiatian Xi and
                  Berrak Sisman and
                  Carlos Busso and
                  Haizhou Li},
  title        = {Towards Emotionally Consistent Text-Based Speech Editing: Introducing EmoCorrector and The {ECD-TSE} Dataset},
  booktitle    = {Proc. Interspeech},
  year         = {2025},
}

@inproceedings{ESD,
  author       = {Kun Zhou and
                  Berrak Sisman and
                  Rui Liu and
                  Haizhou Li},
  title        = {Seen and Unseen Emotional Style Transfer for Voice Conversion with {A} New Emotional Speech Dataset},
  booktitle    = {Proc. {ICASSP}},
  pages        = {920--924},
  year         = {2021},
}

@article{Qwen2.5-Omni,
  author       = {Jin Xu and
                  Zhifang Guo and
                  Jinzheng He and
                  Hangrui Hu and
                  Ting He and others},
  title        = {Qwen2.5-Omni Technical Report},
  journal      = {arXiv preprint arXiv:2503.20215},
  volume       = {abs/2503.20215},
  year         = {2025},
}

@article{qwen3-omni,
  author       = {Jin Xu and
                  Zhifang Guo and
                  Hangrui Hu and
                  Yunfei Chu and
                  Xiong Wang and others},
  title        = {Qwen3-Omni Technical Report},
  journal      = {arXiv preprint arXiv:2509.17765},
  volume       = {abs/2509.17765},
  year         = {2025},
}

@article{c2ser,
  title={Steering language model to stable speech emotion recognition via contextual perception and chain of thought},
  author       = {Zhixian Zhao and
                  Xinfa Zhu and
                  Xinsheng Wang and
                  Shuiyuan Wang and
                  Xuelong Geng and
                  Wenjie Tian and
                  Lei Xie},
  journal={IEEE Transactions on Audio, Speech and Language Processing},
  volume={34},
  pages={415--426},
  year={2025},
  publisher={IEEE}
}

@article{kimi,
  title={Kimi-audio technical report},
  author       = {Ding Ding and
                  Zeqian Ju and
                  Yichong Leng and
                  Songxiang Liu and
                  Tong Liu and
                  Zeyu Shang and
                  Kai Shen and
                  Wei Song and others},
  journal={arXiv preprint arXiv:2504.18425},
  year={2025}
}

@article{step-audio2,
  title={Step-audio 2 technical report},
  author={Wu, Boyong and Yan, Chao and Hu, Chen and Yi, Cheng and Feng, Chengli and Tian, Fei and Shen, Feiyu and Yu, Gang and Zhang, Haoyang and Li, Jingbei and others},
  journal={arXiv preprint arXiv:2507.16632},
  year={2025}
}

@article{glm4,
  title={Glm-4-voice: Towards intelligent and human-like end-to-end spoken chatbot},
  author       = {Aohan Zeng and
                  Zhengxiao Du and
                  Mingdao Liu and
                  Kedong Wang and
                  Shengmin Jiang and
                  Lei Zhao and
                  Yuxiao Dong and
                  Jie Tang},
  journal={arXiv preprint arXiv:2412.02612},
  year={2024}
}

@article{gemini2.5,
  title={Gemini 2.5: Pushing the frontier with advanced reasoning, multimodality, long context, and next generation agentic capabilities},
  author={Comanici, Gheorghe and Bieber, Eric and Schaekermann, Mike and Pasupat, Ice and Sachdeva, Noveen and others},
  journal={arXiv preprint arXiv:2507.06261},
  year={2025}
}

@inproceedings{freeze-omni,
  author       = {Xiong Wang and
                  Yangze Li and
                  Chaoyou Fu and
                  Yike Zhang and
                  Yunhang Shen and
                  Lei Xie and
                  Ke Li and
                  Xing Sun and
                  Long Ma},
  title        = {Freeze-Omni: {A} Smart and Low Latency Speech-to-speech Dialogue Model
                  with Frozen {LLM}},
  booktitle    = {Proc. {ICML}},
  volume       = {267},
  year         = {2025},
}

@article{sapo,
  title={Soft adaptive policy optimization},
  author={Gao, Chang and Zheng, Chujie and Chen, Xiong-Hui and Dang, Kai and Liu, Shixuan and Yu, Bowen and Yang, An and Bai, Shuai and Zhou, Jingren and Lin, Junyang},
  journal={arXiv preprint arXiv:2511.20347},
  year={2025}
}

@article{gpt4o,
  title={Gpt-4o system card},
  author={Hurst, Aaron and Lerer, Adam and Goucher, Adam P and Perelman, Adam and others},
  journal={arXiv preprint arXiv:2410.21276},
  year={2024}
}

@article{qwen2audio,
  title={Qwen2-audio technical report},
  author       = {Yunfei Chu and
                  Jin Xu and
                  Qian Yang and
                  Haojie Wei and
                  Xipin Wei and
                  Zhifang Guo and others},
  journal={arXiv preprint arXiv:2407.10759},
  year={2024}
}

@inproceedings{scene,
  author       = {Guan{-}Ting Lin and
                  Cheng{-}Han Chiang and
                  Hung{-}yi Lee},
  title        = {Advancing Large Language Models to Capture Varied Speaking Styles
                  and Respond Properly in Spoken Conversations},
  booktitle    = {Proc. {ACL}},
  pages        = {6626--6642},
  year         = {2024},
}

@article{index-tts2,
  author       = {Siyi Zhou and
                  Yiquan Zhou and
                  Yi He and
                  Xun Zhou and
                  Jinchao Wang and
                  Wei Deng and
                  Jingchen Shu},
  title        = {IndexTTS2: {A} Breakthrough in Emotionally Expressive and Duration-Controlled
                  Auto-Regressive Zero-Shot Text-to-Speech},
  journal={arXiv preprint arXiv:2506.21619},
  year={2025}
}

@article{PPO,
  author       = {John Schulman and
                  Filip Wolski and
                  Prafulla Dhariwal and
                  Alec Radford and
                  Oleg Klimov},
  title        = {Proximal Policy Optimization Algorithms},
  journal={arXiv preprint arXiv:1707.06347},
  year={2017}
}

@article{GRPO,
  author       = {Zhihong Shao and
                  Peiyi Wang and
                  Qihao Zhu and
                  Runxin Xu and
                  Junxiao Song and
                  Mingchuan Zhang and
                  Y. K. Li and
                  Y. Wu and
                  Daya Guo},
  title        = {DeepSeekMath: Pushing the Limits of Mathematical Reasoning in Open
                  Language Models},
  journal={arXiv preprint arXiv:2402.03300},
  year         = {2024},
}

@inproceedings{salmonn,
  author       = {Changli Tang and
                  Wenyi Yu and
                  Guangzhi Sun and
                  Xianzhao Chen and
                  Tian Tan and
                  Wei Li and
                  Lu Lu and
                  Zejun Ma and
                  Chao Zhang},
  title        = {{SALMONN:} Towards Generic Hearing Abilities for Large Language Models},
  booktitle    = {Proc. {ICLR}},
  year         = {2024},
}

@article{osum,
  author       = {Xuelong Geng and
                  Qijie Shao and
                  Hongfei Xue and
                  Shuiyuan Wang and
                  Hanke Xie and
                  Zhao Guo and
                  Yi Zhao and
                  Guojian Li and
                  Wenjie Tian and
                  Chengyou Wang and
                  Zhixian Zhao and others},
  title        = {OSUM-EChat: Enhancing End-to-End Empathetic Spoken Chatbot via Understanding-Driven
                  Spoken Dialogue},
  journal      = {arXiv preprint arXiv:2508.09600},
  year         = {2025},
}

@article{qwen-audio,
  author       = {Yunfei Chu and
                  Jin Xu and
                  Xiaohuan Zhou and
                  Qian Yang and
                  Shiliang Zhang and
                  Zhijie Yan and
                  Chang Zhou and
                  Jingren Zhou},
  title        = {Qwen-Audio: Advancing Universal Audio Understanding via Unified Large-Scale
                  Audio-Language Models},
  journal      = {arXiv preprint arXiv:2311.07919},
  year         = {2023},
}

@article{emotionthinker,
  author       = {Dingdong Wang and
                  Shujie Liu and
                  Tianhua Zhang and
                  Youjun Chen and
                  Jinyu Li and
                  Helen Meng},
  title        = {EmotionThinker: Prosody-Aware Reinforcement Learning for Explainable
                  Speech Emotion Reasoning},
  journal       = {arXiv preprint arXiv:2601.15668},
  year         = {2026},
}

@inproceedings{e-chat,
  author       = {Hongfei Xue and
                  Yuhao Liang and
                  Bingshen Mu and
                  Shiliang Zhang and
                  Mengzhe Chen and
                  Qian Chen and
                  Lei Xie},
  title        = {E-Chat: Emotion-Sensitive Spoken Dialogue System with Large Language
                  Models},
  booktitle    = {Proc. {ISCSLP}},
  pages        = {586--590},
  year         = {2024},
}

@article{HumDial-EIBench,
  author       = {Shuiyuan Wang and
                  Zhixian Zhao and
                  Hongfei Xue and
                  Chengyou Wang and
                  Shuai Wang and
                  Hui Bu and
                  Xin Xu and
                  Lei Xie},
  title        = {HumDial-EIBench: {A} Human-Recorded Multi-Turn Emotional Intelligence
                  Benchmark for Audio Language Models},
  journal      = {arXiv preprint arXiv: 2604.11594},
  year         = {2026},
}

@article{humdial,
  author       = {Zhixian Zhao and
                  Shuiyuan Wang and
                  Guojian Li and
                  Hongfei Xue and
                  Chengyou Wang and
                  Shuai Wang and
                  Longshuai Xiao and
                  Zihan Zhang and
                  Hui Bu and
                  Xin Xu and
                  Xinsheng Wang and
                  Hexin Liu and
                  Eng Siong Chng and
                  Hung{-}yi Lee and
                  Haizhou Li and
                  Lei Xie},
  title        = {The {ICASSP} 2026 HumDial Challenge: Benchmarking Human-like Spoken
                  Dialogue Systems in the {LLM} Era},
  journal      = {arXiv preprint arXiv: 2601.05564},
  year         = {2026},
}

@inproceedings{emova,
  author       = {Kai Chen and
                  Yunhao Gou and
                  Runhui Huang and
                  Zhili Liu and
                  Daxin Tan and
                  Jing Xu and
                  Chunwei Wang and others},
  title        = {{EMOVA:} Empowering Language Models to See, Hear and Speak with Vivid
                  Emotions},
  booktitle    = {Proc. {CVPR}},
  pages        = {5455--5466},
  year         = {2025},
}

@article{deepseek-v3,
  author       = {DeepSeek{-}AI},
  title        = {DeepSeek-V3 Technical Report},
  journal      = {arXiv preprint arXiv:2412.19437},
  year         = {2024},
}

@article{deepseekr1,
  title={DeepSeek-R1: Incentivizing Reasoning Capability in {LLMs} via Reinforcement Learning},
  author={DeepSeek-AI},
  journal={arXiv preprint arXiv:2501.12948},
  year={2025}
}

@article{SABER,
  author       = {Zhixian Zhao and
                  Wenjie Tian and
                  Xiaohai Tian and
                  Jun Zhang and
                  Lei Xie},
  title        = {Integrating Fine-Grained Audio-Visual Evidence for Robust Multimodal
                  Emotion Reasoning},
  journal       = {arXiv preprint arXiv:2601.18321},
  year         = {2026},
}

@article{emoomni,
  title={{EmoOmni}: Bridging Emotional Understanding and Expression in Omni-Modal {LLMs}},
  author={Wenjie Tian and Zhixian Zhao and Jingbin Hu and Huakang Chen and Haohe Liu and Binshen Mu and Lei Xie},
  journal={arXiv preprint arXiv:2602.21900},
  year={2026}
}

@article{text1,
  title={Audiopalm: A large language model that can speak and listen},
  author       = {Paul K. Rubenstein and
                  Chulayuth Asawaroengchai and
                  Duc Dung Nguyen and
                  Ankur Bapna and
                  Zal{\'{a}}n Borsos and others},
  journal={arXiv preprint arXiv:2306.12925},
  year={2023}
}

@inproceedings{text2,
  title={Listen, Think, and Understand},
  author       = {Yuan Gong and
                  Hongyin Luo and
                  Alexander H. Liu and
                  Leonid Karlinsky and
                  James R. Glass},
  booktitle={Proc. {ICLR}},
  year={2024}
}

@article{conflict1,
  author       = {Dawei Huang and
                  Yongjie Lv and
                  Ruijie Xiong and
                  Chunxiang Jin and
                  Xiaojiang Peng},
  title        = {When Tone and Words Disagree: Towards Robust Speech Emotion Recognition
                  under Acoustic-Semantic Conflict},
  journal       = {arXiv preprint arXiv:2601.04564},
  year         = {2026},
}

@article{conflict2,
  author       = {Jingyi Chen and
                  Zhimeng Guo and
                  Jiyun Chun and
                  Pichao Wang and
                  Andrew Perrault and
                  Micha Elsner},
  title        = {Do Audio LLMs Really LISTEN, or Just Transcribe? Measuring Lexical
                  vs. Acoustic Emotion Cues Reliance},
  journal       = {arXiv preprint arXiv:2510.10444},
  year         = {2025},
}

\end{document}